\documentclass{pasa}%

\usepackage{graphicx}

\title[GOTO with the LSST stack {\sc i}: coadds]{Processing GOTO data with the Rubin Observatory LSST Science Pipelines {\sc i} : Production of coadded frames}

\def\CC{{C\nolinebreak[4]\hspace{-.05em}\raisebox{.4ex}{\tiny\bf ++}}}

\author[J. R. Mullaney et al.]{
J. R. Mullaney,$^{1}$ L. Makrygianni,$^{1}$ V. Dhillon,$^{1}$ S.
Littlefair,$^{1}$ K. Ackley,$^{2}$ M. Dyer,$^{1}$ J. Lyman,$^{3}$ K.
Ulaczyk,$^{3}$ R. Cutter,$^{3}$ Y.-L. Mong,$^{2}$ D. Steeghs,$^{3}$ D.
K. Galloway,$^{2,4}$ P. O'Brien,$^{5}$ G. Ramsay,$^{6}$ S.
Poshyachinda,$^{7}$ R. Kotak,$^{8}$ L. Nuttall,$^{9}$ E.
Pall\'e,$^{10}$ D. Pollacco,$^{3}$ E. Thrane,$^{2}$ S.
Aukkaravittayapun,$^{7}$ S. Awiphan,$^{7}$ R. Breton,$^{11}$ U.
Burhanudin,$^{1}$ P. Chote,$^{3}$ A. Chrimes,$^{3}$ E. Daw,$^{1}$ C.
Duffy,$^{6}$ R. Eyles-Ferris,$^{5}$ B. Gompertz,$^{3}$ T.
Heikkil\"a,$^{8}$ P. Irawati,$^{7}$ M. Kennedy,$^{11}$ T.
Killestein,$^{3}$ A. Levan,$^{3}$ T. Marsh,$^{3}$ D. Mata-
Sanchez,$^{11}$ S. Mattila,$^{8}$ J. Maund,$^{1}$ J. McCormac,$^{3}$
D. Mkrtichian,$^{7}$ E. Rol,$^{2}$ U. Sawangwit,$^{7}$ E.
Stanway,$^{3}$ R. Starling,$^{5}$ S. Tooke,$^{5}$ K. Wiersema,$^{3}$
\affil{$^{1}$Department of Physics and Astronomy, University of Sheffield, Sheffield S3 7RH, UK}
\affil{$^{2}$School of Physics \& Astronomy, Monash University, Clayton VIC 3800, Australia}
\affil{$^{3}$Department of Physics, University of Warwick, Gibbet Hill Road, Coventry CV4 7AL, UK}
\affil{$^{4}$OzGrav: The ARC Centre of Excellence for Gravitational Wave Discovery, Clayton VIC 3800, Australia}
\affil{$^{5}$School of Physics \& Astronomy, University of Leicester, University Road, Leicester LE1 7RH, UK}
\affil{$^{6}$Armagh Observatory \& Planetarium, College Hill, Armagh, BT61 9DG}
\affil{$^{7}$National Astronomical Research Institute of Thailand, 260 Moo 4, T. Donkaew, A. Maerim, Chiangmai, 50180 Thailand}
\affil{$^{8}$Department of Physics \& Astronomy, University of Turku, Vesilinnantie 5, Turku, FI-20014, Finland}
\affil{$^{9}$University of Portsmouth, Portsmouth, PO1 3FX, UK}
\affil{$^{10}$Instituto de Astrof'{i}sica de Canarias, E-38205 La Laguna, Tenerife, Spain}
\affil{$^{11}$Jodrell Bank Centre for Astrophysics, Department of Physics and Astronomy, The University of Manchester, Manchester M13 9PL, UK}
}
\jid{PASA}
\doi{10.1017/pas.\the\year.xxx}
\jyear{\the\year}

\usepackage{aas_macros}
\usepackage{hyperref} 
\hypersetup{colorlinks,citecolor=blue,linkcolor=blue,urlcolor=blue}

\hypersetup{draft}
\begin{document}

\begin{frontmatter}
\maketitle

\begin{abstract}
The past few decades have seen the burgeoning of wide field, high cadence surveys, the most formidable of which will be the Legacy Survey of Space and Time (LSST) to be conducted by the {Vera C. Rubin Observatory}. So new is the field of systematic time-domain survey astronomy, however, that major scientific insights will continue to be obtained using smaller, more flexible systems than the LSST. One such example is the Gravitational-wave Optical Transient Observer (GOTO), whose primary science objective is the optical follow-up of Gravitational Wave events. The amount and rate of data production by GOTO and other wide-area, high-cadence surveys presents a significant challenge to data processing pipelines which need to operate in near real-time to fully exploit the time-domain. In this study, we adapt the {Rubin Observatory LSST Science Pipelines} to process GOTO data, thereby exploring the feasibility of using this ``off-the-shelf'' pipeline to process data from other wide-area, high-cadence surveys. In this paper, we describe how we use the { LSST Science Pipelines} to process raw GOTO frames to ultimately produce calibrated coadded images and photometric source catalogues. After comparing the measured astrometry and photometry to those of matched sources from PanSTARRS DR1, we find that measured source positions are typically accurate to sub-pixel levels, and that measured L-band photometries are accurate to $\sim50~${\rm mmag} at $m_L\sim16$ and $\sim200~${\rm mmag} at $m_L\sim18$. { These values compare favourably to those obtained using GOTO's primary, in-house pipeline, {\sc gotophoto}, in spite of both pipelines having undergone further development and improvement beyond the implementations used in this study.} { Finally, we release a generic ``obs package'' that others can build-upon should they wish to use the LSST Science Pipelines to process data from other facilities.}
\end{abstract}

\begin{keywords}
Astronomy data analysis -- Surveys -- Astrometry -- Photometry
\end{keywords}
\end{frontmatter}

\section{INTRODUCTION }
\label{sec:intro}

Since the undertaking of the National Geographic Society -- Palomar Observatory Sky Survey (NGS--POSS) during the 1940's and 1950's (\citealt{Abell59,Minkowski63}), wide-area surveys have played an increasingly important role within astronomy research. Such are their importance that wide-area surveys have been conducted in bands spanning the whole of the observable electromagnetic spectrum, from radio through to gamma rays (see \citealt{Lawrence07,Djorgovski13} for reviews). Usually such surveys are commissioned with a handful of primary scientific goals in mind, such as measuring the large scale structure of the Universe (in the case of the Sloan Digital Sky Survey, or SDSS; \citealt{York00}) or the nature of Dark Energy (the Dark Energy Survey, or DES; \citealt{DESC05}). In most cases, however, their scientific impact is ultimately recognised as extending far beyond their original remit, not least in the discovery of new science or classes of object that warrant further study.

Early wide-area optical surveys such as the NGS-POSS and its southern counterpart, the ESO/SERC (\citealt{Holmberg74}), conducted a single pass of the sky, often in multiple filters or bandpasses in order to obtain colour or crude spectral information (i.e., spectral indices). These single-pass surveys provided astronomers with a ``static'' view of the Universe. Despite many diverse areas of astronomy benefiting from such static surveys, they are unable to provide much material information on time-varying or transient processes (aside from being used as a reference against which later unrelated surveys or pointed observations are compared; e.g., \citealt{Ross18}). The past two decades, however, have seen significant investments in wide-area ``time-domain'' surveys which conduct repeated passes of the sky, thereby enabling the study of how astronomical objects change over time (e.g., OGLE, \citealt{Udalski97}; SuperWASP, \citealt{Pollacco06}; Catalina Sky Survey, \cite{Drake09}; PanSTARRS, \citealt{Chambers16}; ZTF, \citealt{Bellm19}). In some cases such as PanSTARRS (\citealt{Chambers16}), optical time-domain surveys have been (at least partly) motivated by the desire to identify near-Earth objects which would lead to extinction-level events should they impact the Earth. However, as with static wide-field surveys, such time-domain surveys have been exploited to gain insights into a wide range of other phenomena, including supernovae, exoplanets, microlensing, and AGN variability, etc.

The most ambitious wide-area time-domain survey is that which will be conducted by the { Vera C. Rubin Observatory} (\citealt{Ivezic19}) which, at the the time of writing, is under construction on the summit of Cerro Pach\'{o}n in the Chilean Andes. The { Rubin Observatory} hosts an 8~m-aperture telescope that will repeatedly survey the sky in six wavebands. The resulting survey -- referred to as the Legacy Survey of Space and Time (LSST) - will reach a single-pass r-band depth of around 24.5 magnitudes, but will ultimately reach a r-band depth of around 27.5 magnitudes across the observable sky on the coaddition of multiple passes. The Rubin Observatory's camera will consist of 189 16~megapixel CCDs, each of which will deliver, on average, 1000 science frames per night (corresponding to around 200,000 science exposures, representing around 20~TB of data, per night). The Rubin Observatory is developing its own pipeline (the { LSST Science Pipelines}, hereafter the LSST stack; \citealt{Juric17, Juric19, Bosch19}) that is capable of processing this data at the required rate. As well as ``standard'' optical astronomy data processing steps (i.e., calibration, background subtraction, source detection and measuring), the LSST stack must also perform additional processing steps to fully exploit the time-domain aspect of the survey. These include the coaddition of multiple epochs of data, ``forced'' photometry (in which the properties of a source detected in a reference image are measured in a new exposure, irrespective of whether it detected in the latter), and difference imaging (whereby a reference image is subtracted from a new exposure in order to more easily identify transient sources or sources that have varied between exposures). As such, as well as breaking new ground in terms of telescope technology, the LSST also represents an ambitious software project.

With much still to learn about the time-varying nature of astronomical sources, significant scientific insights can be gained by projects with far less resources than the Rubin Observatory. For example, the SuperWASP (\citealt{Pollacco06}) and ASASN (\citealt{Shappee14}) projects have conducted groundbreaking science with hardware that is within financial reach of smaller collaborations of research institutes. The relative ease of deployment and flexibility of such facilities mean they will continue to play an important role in time-domain astronomy even during the LSST era. One such example is the Gravitational Wave Optical Observer (GOTO; Steeghs et al; in prep.; see section \ref{sec:goto}), whose primary science objective is to identify the optical counterparts of Gravitational Wave (GW) events by quickly scanning the localisation skymap provided by GW detectors (currently LIGO and VIRGO). To identify the optical counterparts, these scans must be compared against recently-obtained reference images which are obtained through repeated surveys conducted during times when GOTO is not following-up trigger events. GOTO's ``survey-mode'' thereby represents a high-cadence (i.e., daily to weekly), wide-area time-domain survey similar to, if somewhat shallower than, the LSST.

The GOTO collaboration is developing its own dedicated in-house pipeline -- {\sc gotophoto} -- as its primary data processing system. However, the conceptual similarity of the GOTO survey to that of the LSST make the LSST stack a viable secondary means to process GOTO data, which is useful for cross-comparison and verification purposes. Indeed, the LSST stack has been designed from the outset to be able to process data from other facilities (see \citealt{Bosch18} for such an example). In this regard, the GOTO survey provides a ``real-world'' time domain survey testbed for the LSST stack in addition to simulated LSST data. With these points in mind, we have successfully implemented the LSST stack to process GOTO data in near real-time. The aim of this paper is to outline the steps we took to achieve this goal up to the point of producing coadded images.\footnote{A second paper in this series, Makrygianni et al. (2020), outlines the steps taken to perform forced photometry on GOTO data, using the sources detected in the coadds as references.} In the process, we also wrote our own additional modules that call upon standard LSST modules to address our specific needs. This latter point further emphasises the flexibility of the LSST stack as a general software suite to process other wide-area survey data. With this in mind, it is our intention that the methodology, software, and quality assurance checks laid-out in this paper can be used as additional resources for other future wide-area surveys considering using the LSST stack as a viable data processing pipeline. { It should be noted, however, that both the LSST stack and {\sc gotophoto} are still under active development, and some of the features and steps described in this paper have either become or are close to becoming obsolete in the latest versions of the LSST stack, particularly those related to astrometric calibration and deblending.}

\begin{figure}
    \includegraphics[width=\linewidth]{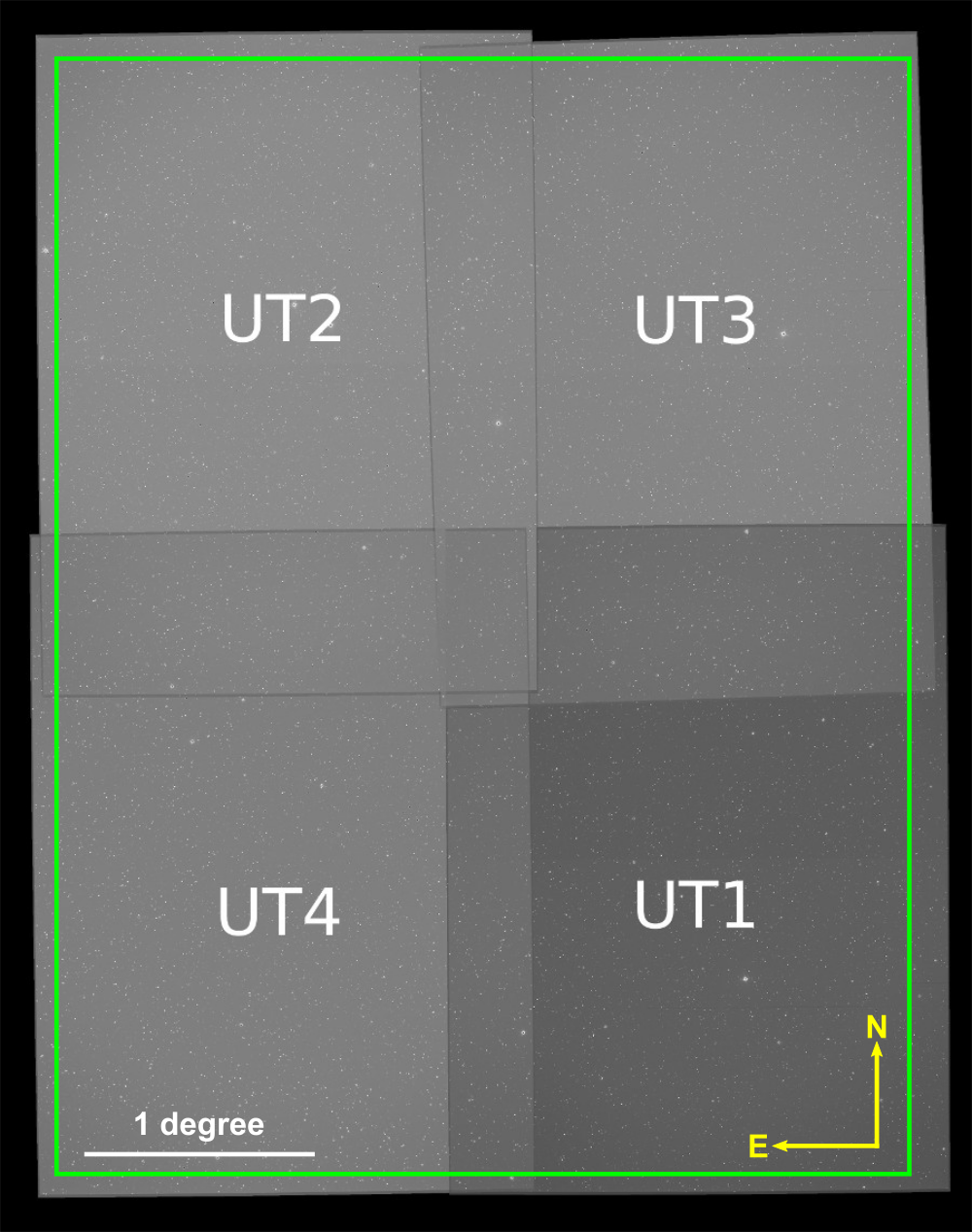}
    \caption{The field of view of the four UTs that were installed on the GOTO mount during the prototype phase when the data used in this study was obtained. The orientation is shown in the bottom right, while the scale is shown in the bottom left of the image. The green box indicates the size of a single GOTO ``tile'' during prototype mode; these tiles are used split up the celestial sphere into an easily indexable grid for scheduling purposes.}
    \label{fig:fov}
\end{figure}
 
This paper is structured as follows: the following section describes GOTO in more detail, with a particular emphasis on its data products. Section \ref{sec:lsst} provides a brief overview of the LSST stack, while section \ref{sec:obs} describes the steps we took to enable the LSST stack to process GOTO data. In section \ref{sec:proc} we describe the act of processing the raw data (i.e., the various tasks called and their outputs), before providing a brief conclusion in section \ref{sec:conc}. Finally, in an Appendix, we describe a publicly-available generic  ``obs package'' (see Section \ref{sec:lsst}) that can be used as a starting point for others wishing to utilise the LSST stack to process their data.

\section{The GOTO survey}
\label{sec:goto}
As described in the Introduction, our aim at the start of this study was to process data obtained by GOTO by using the LSST stack. In this section, we provide a brief description of the GOTO observatory and data products in order to provide context for later sections. A more detailed description of the GOTO observatory and its control systems can be found in Steeghs et al. (in prep.) and \cite{Dyer18}, respectively.

\begin{figure}
    \includegraphics[width=\linewidth]{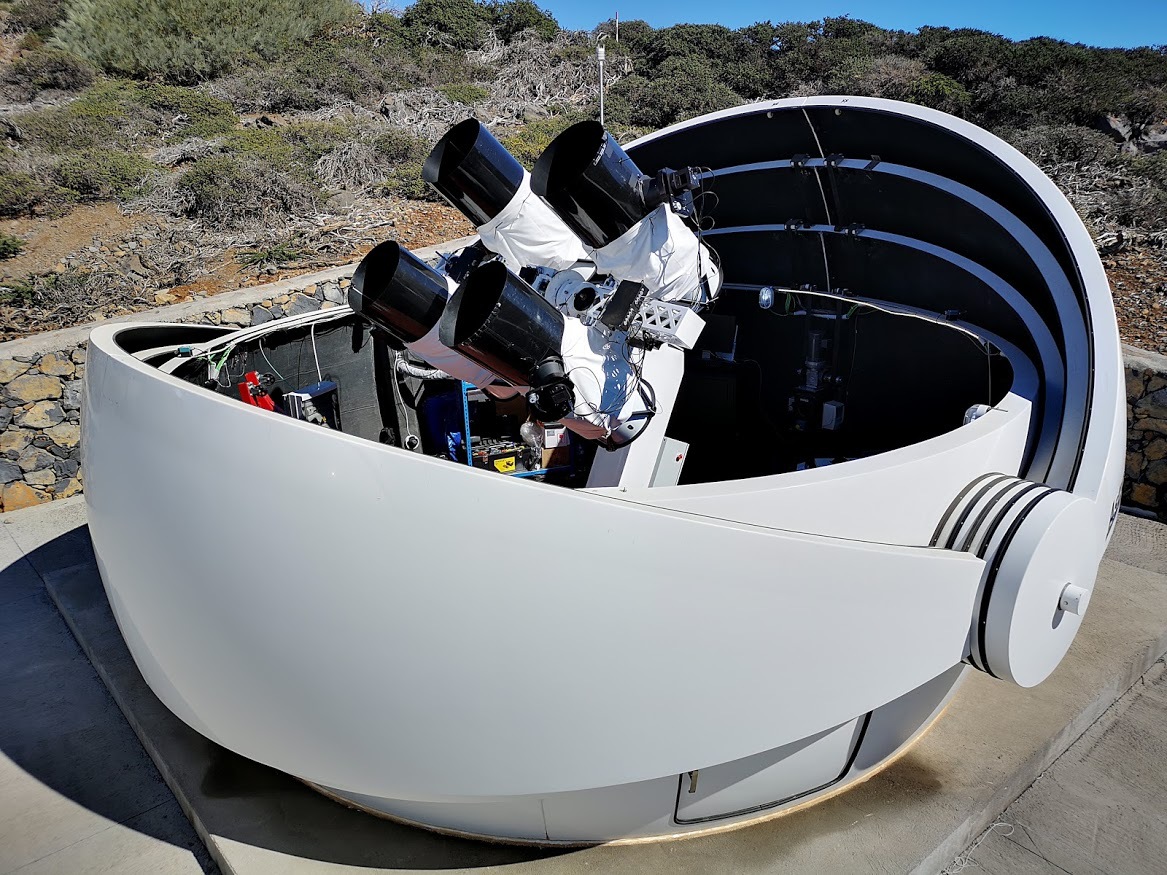}
    \caption{A photograph of the first GOTO mount located on La Palma during its prototype phase when it was equipped with four UTs. Since this photograph was taken four more UTs have been added to this mount. Image from \protect \cite{Dyer20}.}
    \label{fig:goto}
\end{figure}

\subsection{The GOTO observatory}
GOTO's primary scientific goal is to survey the localisation skymaps of gravitational wave events provided by ground-based gravitational-wave detectors. At the time of writing, the three operational gravitational wave detectors (consisting of the two LIGO detectors and the single VIRGO detectors; \citealt{Aasi15} and \citealt{Acernese15}, respectively) announce gravitational wave events with localisation skymaps typically covering an on-sky area of a few hundred square degrees (e.g., \citealt{Fairhurst11, Grover14}). Since any optical counterpart to a GW event is likely to fade rapidly, it is important that GOTO is able to scan the localisation skymaps quickly in order to both maximise the likelihood of detection (i.e., before the counterpart fades to below GOTO's sensitivity) and to minimise the time taken to pass accurate positional information to other telescopes for more detailed follow-up observations. Such rapid scanning of the large GW localisation skymap requires an observatory with a comparatively large field-of-view. To achieve this, GOTO adopts a modular design, consisting of eight individual telescopes (hereafter ``Unit Telescopes'', or UTs), each with a roughly $2.8 \times 2.1\approx6$~sq. degree field-of-view,  attached to a single mount. Each UT is equipped with a single 50 megapixel CCD with a plate scale of 1.24 arcsec per pixel. The UTs are aligned such that their fields-of-view are slightly offset from each other, forming a contiguous field-of-view of roughly 40 sq. degrees per mount pointing after accounting for overlap and vignetting (see Fig. \ref{fig:fov}). We note that, aside from manual alignment, the UTs cannot be moved independently of each other. Each UT has its own filter wheel containing the Baader R, G, and B filters, plus a fourth Baader L-band filter. This latter filter is a broad-band filter that covers the entire optical spectrum between roughly 4000~\AA\ and 7000~\AA\ (i.e., roughly the G and R bands) and is used to maximise the amount of light that reaches the detectors. Under normal circumstances, GOTO is operated remotely as a fully-robotic observatory using a control system described in \cite{Dyer18}.

At the time of writing, GOTO consists of a single mount that is equipped with eight UTs and is based at the Observatorio del Roque de los Muchachos on La Palma, Spain. GOTO's full design specification -- for which funding has been secured -- includes a second, eight-UT mount on La Palma (completing GOTO-North) and two further eight-UT mounts based at Siding Springs, Australia (i.e., GOTO-South). The research described in this paper, however, is based on data acquired when GOTO was in its prototype phase, when it consisted of single mount equipped with four UTs (\citealt{Gompertz20}; see Fig. \ref{fig:goto}).

\begin{figure}
    \includegraphics[width=\linewidth]{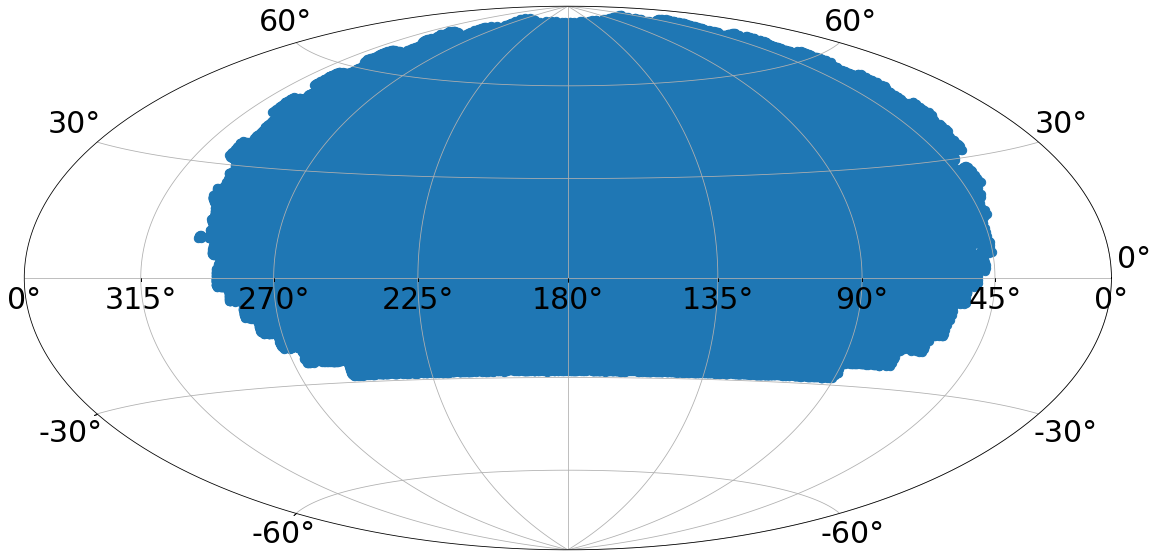}
    \caption{The coverage of the GOTO observations we have processed by the LSST stack (shown in blue). These observations were made between 2019-02-24 and 2019-03-12 and avoid the densest regions of the Galactic plane (see section \ref{sec:gotoData})}
    \label{fig:coverage}
\end{figure}

\subsection{GOTO data}
\label{sec:gotoData}
GOTO operates under two main observing modes. The first mode is associated with transient follow-up, which involves scanning the sky for the afterglow of GW events or other triggers such as gamma ray bursts. The second mode consists of survey observations, during which an archive of images is built up, primarily to be used as reference images against which transient follow-up images can be compared to identify new sources (i.e., potential optical counterparts to the aforementioned triggers). This second mode also represents a time-domain survey which can be used for a multitude of other science objectives (e.g., supernovae searches, stellar flares, AGN variability, etc.). Irrespective of mode, GOTO almost always observes on a fixed grid of ``tiles'' to ensure that the same part of the sky is always covered by roughly the same region of the same UT.\footnote{GOTO can be operated ``off-grid'', but this feature is rarely used in practice.} Most survey observations are conducted with the L-band filter, although less frequent surveys are conducted with the R, G, and B filters, primarily to aid source classification.

While operating in its survey mode, each of GOTO's operating UTs take three one-minute exposures. As such, when operating with a full cohort of eight UTs, GOTO currently delivers 24 exposures per pointing. With a nominal pointing lasting four minutes (including overheads), this corresponds to roughly 3600 exposures per (10-hour) night, corresponding to 150 pointings covering, in total, roughly $1.2\times10^{4}$ square degrees. The need for an automated, highly stable pipeline to process this data is, therefore, clear. To address this need, the GOTO collaboration has developed its own in-house data processing pipeline -- {\sc gotophoto}. While {\sc gotophoto} is used as GOTO's primary data processing pipeline, we also felt it would be worthwhile to assess the viability of other pipelines, including the LSST stack.

The data used throughout this study were obtained by GOTO between 2019-02-24 and 2019-03-12 (inclusive), while it was observing in survey mode. The data covers the region of the sky spanning ${\rm 2\ h\lesssim RA\lesssim20\ h}$ and ${\rm -20\deg\lesssim Dec\lesssim90\deg}$, which avoids the densest parts of the Galactic plane (see fig. \ref{fig:coverage}). While GOTO observed the full northern sky during its prototype phase, we chose not to process the observations covering the Galactic plane as our interests are in using the LSST stack-produced results for extragalactic science.\footnote{By contrast, {\sc gotophoto} has been used to process all GOTO observations, including those covering the Galactic plane.} Typically, each tile that goes into our coadds has been observed once, with each observation consisting of three back-to-back 1-minute exposures.

\section{The LSST Science Pipelines}
\label{sec:lsst}
In many respects, GOTO's survey mode resembles the wide-area survey that will be carried out by the LSST: a large area of sky surveyed multiple times per year with a wide-area telescope consisting of multiple CCDs. As such, the software being developed to process LSST data performs many of the tasks needed to also process GOTO data. Thankfully, the LSST stack has been designed in such a way that it can be used to process data produced by facilities other than the LSST. Indeed, the LSST stack is the primary processing pipeline being used to process data taken with HyperSupremeCam on Subaru (\citealt{Bosch18}). In this section, we provide a brief overview of the LSST stack to provide context for the next section in which we describe how we adapted the LSST stack to process GOTO data.

The LSST stack is written in a combination of the Python and \CC\ programming languages. The latter is used for the bulk of the calculations for reasons of improved performance, but knowledge of \CC\ is not required to use the LSST stack. The entirety of the LSST stack is published under the GNU General Public Licence at \url{https://github.com/lsst}. We are able to use the LSST stack to conduct the following processing steps of our GOTO data: 

\begin{itemize}
\item {\bf Data ingestion}, in which information (such as exposure type $[$science, flat, bias, etc.$]$, filter name, CCD number, date of observation, etc.) is extracted from the FITS file headers and used to populate a database of observations. Files are also either copied or linked from their original locations to new locations which satisfy a standard naming convention;
\item {\bf Construction of master calibration frames}, in which flats, bias, and dark frames from multiple nights are median-combined and ingested;
\item {\bf Instrument signature removal (ISR)}, whereby science frames are corrected using the aforementioned master bias, dark, and flat frames;
\item {\bf Cosmic ray identification} and removal;
\item {\bf Background subtraction} using a low-order 2D polynomial fit to the background;
\item {\bf Modelling of the Point Spread Function (PSF)}, allowing for PSF variance across the science frame;
\item {\bf Astrometric and photometric calibration} by comparison to external reference catalogues;
\item {\bf Source detection, deblending, and measurement} on single science frames;
\item {\bf Frame alignment and coaddition} to obtain deep coadded science images followed by source detection, deblending, and measurement on coadded frames;
\item {\bf Forced photometry} on new frames using the sources detected in the coadded frames as references.
\item {\bf Difference imaging}, whereby a reference image is subtracted from an incoming science image to identify sources that have changed in the intervening period.
\end{itemize}

In this paper, we will describe how we have used the LSST stack to process GOTO data up to frame coaddition. Subsequent papers will explore forced photometry (Makrygianni et al. 2020) and difference imaging (which use the coadded frames and resulting catalogues as references).

\subsection{Obtaining and running the LSST Science Pipelines}
\label{sec:lsst:run}
In order to utilise the LSST stack to process GOTO data, we installed the software and its prerequisites on a local machine. There are currently three ways to obtain and run the LSST stack:
\begin{itemize}
    \item installing locally from source using {\tt lsstsw} and {\tt lsst-build};
    \item installing locally using {\tt newinstall.sh} which creates a self-contained environment from which you can run the lsst stack;
    \item download (and, if required, modify) the LSST {\tt Docker} image and run it as a container.
\end{itemize}
We chose the latter, as it is fully platform-independent and, with the creation of a Dockerfile, easily allows us to build-upon the LSST {\tt Docker} image to include the additional packages required to process GOTO data. A version of our Dockerfile is maintained on \url{https://github.com/jrmullaney/lsstDocker}. We used version 17.0 of the LSST stack which was the stable release when processing our data in March, 2019. As such, the following descriptions of the configuration and modification of the LSST stack is relevant to that version. All our data processing was performed on a Dual Intel Xeon E5-2697v3 2.60Ghz CPU with 28 cores/56 threads with access to 256~GB of RAM. Throughout, we report the wall-clock time to process a single image (exposure or patch; see section \ref{subsec:imageCoaddition}) on a single core, mean-averaged over 10 images.\footnote{Our reported processing times are only to be used as a rough guide, as actual processing times depend heavily on factors such as the number of detected sources in an image and the number and complexity of measured source properties.} It is important to note, however, that most processing steps are embarrassingly parallel across images (i.e., each image can be processed in isolation with little-to-no communication between processors needed). We used the LSST stack's built-in scheduler (\texttt{ctrl\_pool}) to distribute tasks across multiple threads, thereby dramatically reducing total wall-clock processing times.

\section{The \texttt{obs\_goto} package}
\label{sec:obs}
Once installed, we configured the LSST stack to process GOTO data. Configuration of the LSST stack for a specific telescope/camera is achieved by developing a so-called ``obs package''; in our case, \texttt{obs\_goto}. An obs package contains all the information specific to a given camera that the LSST stack requires to process that camera's data. { It is important to note that \texttt{obs\_goto} utilises the now near-obsolete ``Generation 2'' Butler to organise and retrieve data, which, in turn, was built around the \texttt{daf\_persistence} framework. Under that model}, a basic obs package consisted of five main components: a set of files to configure the processing steps, a policy file which provides the on-disk locations and formats of input and output data, a set of python scripts which allow further configuration and the bypassing/modification of default processes, a description of the detector, and a list of packages that must be set up to process the data. Our \texttt{obs\_goto} package is available online via GOTO's project github pages.\footnote{\url{https://github.com/GOTO-OBS/obs_goto}} { The imminent retirement of the Generation 2 Butler in preference of the Generation 3 Butler (\citealt{Jenness19}) means \texttt{obs\_goto} will need to undergo a major re-write to ensure continued compatibility.}

\section{Processing GOTO data with the LSST Science Pipelines}
\label{sec:proc}
In this section we outline the steps we take to process GOTO data using the LSST stack. This begins with data ingestion, and ends with a set of output catalogue files that can be, for example, ingested into a suitable database system (e.g., MySQL, PostgreSQL, etc.). { Throughout this section we name the command-line tasks that we executed to conduct each processing step. Again, it is important to highlight that these are Generation 2 executables; task execution follows a different model under the new Generation 3 framework.}

\subsection{Data ingestion}
Prior to processing any data, we first ingest our images. This step is performed by the \texttt{ingestImages.py} command, which takes the file path to the raw image data as an input parameter.  During the ingest step, the LSST stack extracts specific image metadata (from the image header) and uses it to populate a database with information that allows data to be easily found further downstream. In our case, we populate this database with information related to the type of image data (in our case, bias, dark, flat, or science data), exposure time, filter, ccd identification number, unique observation identification number, and date of observation. This allows us to select for processing those images associated with a given date range, for example, and the stack will refer to the database to identify all other necessary information to uniquely identify the requested frames.

While ingesting, the LSST stack will also rename (via copying, moving or soft-linking to the original) the image data so that it conforms to the naming convention outlined in the policy file (see section \ref{sec:obs}) \footnote{ This has also changed under the Generation 3 Butler with the ingested filename convention no longer described in the policy file}. In our case, we choose to soft-link to the raw data as it avoids unnecessary data duplication and means the raw data can still be accessed via the original file path, should it be necessary.

\subsection{Construction of master calibration frames}
\label{mastercals}
With the raw data ingested, we next generated a set of master calibration (i.e., bias, dark, flat) frames. This is performed using the \texttt{constructBias.py}, \texttt{constructDark.py}, and \texttt{constructFlat.py} commands, respectively. { Since during the ingest stage we requested that the type of each raw frame (e.g., dark, bias, flat, science)} was incorporated into the image database, we specified the data type as an input parameter to these commands, thereby relying on the LSST stack to refer the the image database to identify all appropriate input images to generate the master bias, dark, and flat frames. We produced nightly master bias, dark, and flat frames after manually checking for and, if necessary, removing any low-quality calibration frames (e.g., flats with low numbers of counts). Each of our nightly master calibration frames was generated by median-combining the respective input images.

Following the construction of the each master calibration, they themselves must be ingested using the \texttt{ingestCalibs.py} command. This performs a broadly similar role as the \texttt{ingestImages.py} command described above, but which has been written specifically for the ingestion of calibration frames, and allows for a validity time window to be specified to ensure that a given calibration frame is only used to correct science frames that are taken within a given number of days of the calibration frames. 

\subsection{Processing individual science images}
Having produced and ingested a set of master calibration frames, we were able to start the processing of individual science frames with the LSST stack. This is performed by the \texttt{processCcd.py} command, or its multi-node equivalent \texttt{singleFrameDriver.py}. These scripts perform a number of tasks, starting with instrument signature removal (ISR; i.e., correction for master bias, dark and flat, in our case), followed by image characterisation (involving background subtraction, PSF measurement, and cosmic-ray removal) and image calibration (involving astrometric and photometric calibration). In our case, both image characterisation and calibration required considerable configuration beyond the default settings, which we describe below.

\subsubsection{Instrument Signature Removal}
 To perform ISR, \texttt{singleFrameDriver.py} uses the master bias, darks, and flats described in Section \ref{mastercals}. In the case of the processing the science frames that go into generating our coadds, each frame is corrected using the same night's master calibration frames. At this stage, we do not deviate significantly from the default parameters used by the LSST stack, although we choose not to save ISR-corrected images in order to reduce disk usage. We also turn-off the option to interpolate saturated pixels in the science exposures, but note that saturated pixels are flagged, with the flag mask referred-to in later steps. 

\subsubsection{Image Characterisation}
During image characterisation, the LSST stack first models and subtracts an estimate of the spatially-varying background light. We found that the LSST stack's default parameters that control the background model produced satisfactory results for our GOTO images, so we retained those defaults.

\begin{figure}
    \includegraphics[width=\linewidth]{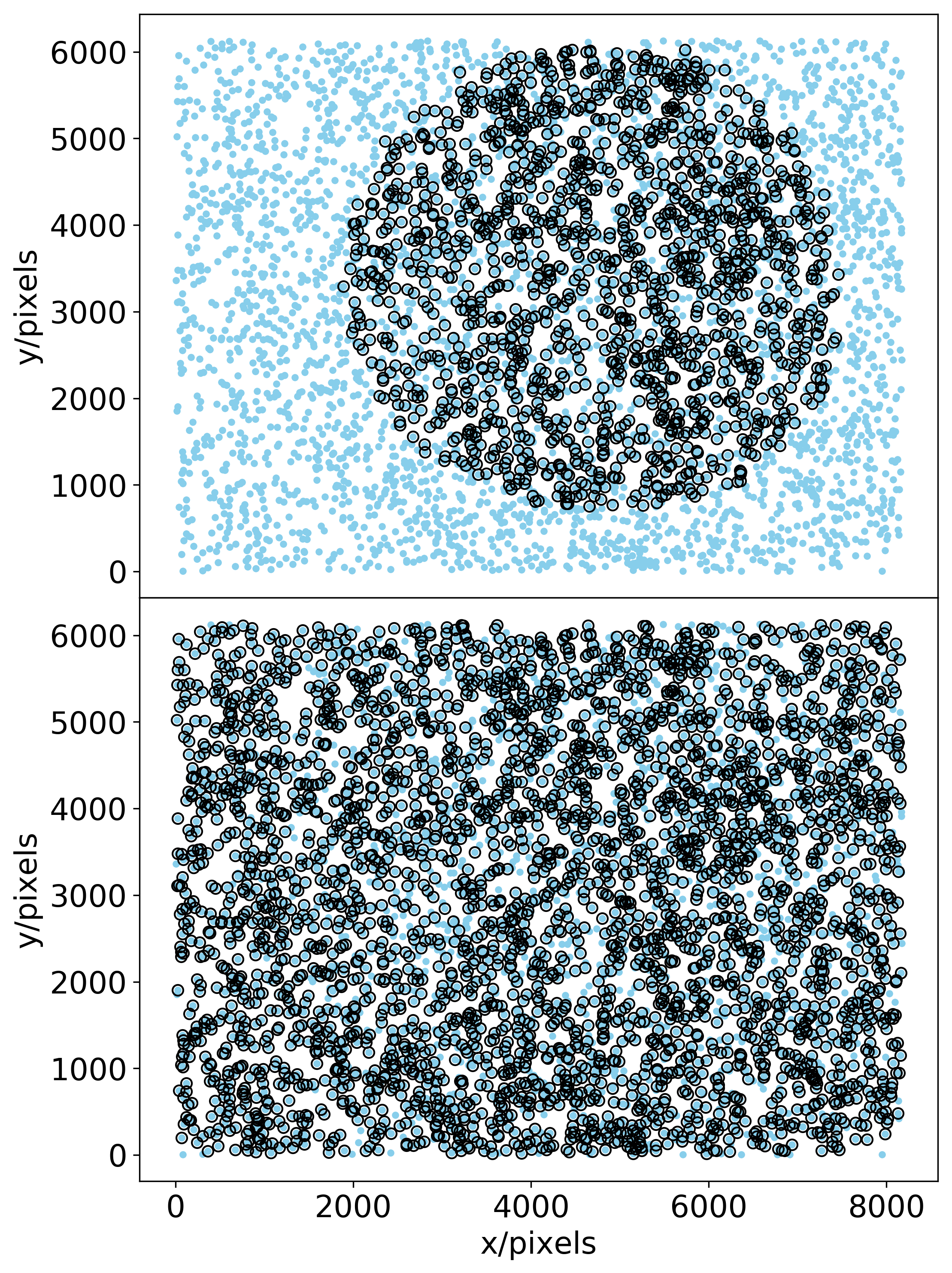}
    \caption{Plots showing the pixel positions of the sources that the LSST stack used to construct a model of the spatially-varying PSF within an individual GOTO frame. The light blue points show the positions of all sources detected at $>100\sigma$ within the frame, whereas the black circles represent those that have been selected as candidates for PSF modelling based on their shape and size. The top panel shows the selection resulting from the LSST stack's default selection criteria, whereas the bottom panel shows the selection after we relaxed these criteria to select sources spanning a wider range of shapes and sizes across the image. Prior to relaxing the selection criteria, large areas of the frame were neglected by the source selector, meaning the PSF model was poorly constrained within the outskirts of the image.}
    \label{fig:psfSelection}
\end{figure}

Following background subtraction, the LSST stack attempts to measure the (spatially-varying) PSF across the image. PSF measurement requires a sample of point-sources to be identified. Ideally, these point sources will be distributed across the full science image to enable the variation of the PSF across the image plane to be modelled. We use a detection threshold of $100\sigma$ to identify an initial sample of robust sources for PSF characterisation; as a single GOTO frame covers a large area, we always have sufficient numbers of bright stars across the whole field-of-view that meet this detection criterion.

\begin{figure*}
    \includegraphics[width=\linewidth]{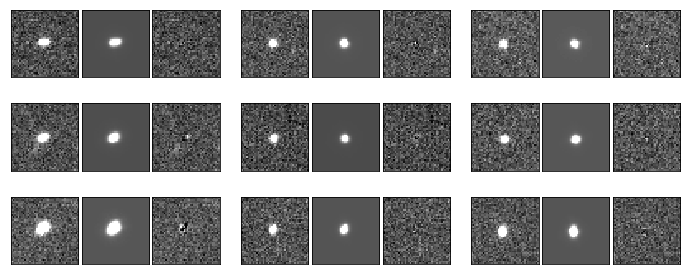}
    \caption{Images showing the spatially-varying PSF at nine different positions across an individual GOTO frame (leftmost image of each group of three images). Also shown is the PSF model at the same position (central image of each group) and the residuals remaining after subtracting the normalised PSF model from the source (rightmost image of each group). All images in each group are scaled equally in brightness. Each group of three images is taken from different regions of the frame, with the upper left group taken from the upper left region of the frame etc. In most cases, the residuals are indistinguishable from noise, which demonstrates the model's ability to reproduce the complexity of GOTO's spatially-varying PSF.} 
    \label{fig:psfs}
\end{figure*}

At this stage, the initial sample of sources includes both point (e.g., bright stars and quasars) and extended (e.g., resolved galaxies) sources, the latter of which need to be removed to ensure only point sources are used for PSF characterisation. For this, we use the default star-selector which uses the size of an object to determine whether it is a star (or, more generally, an unresolved point source) or not. This is achieved by using a K-means algorithm to identify a cluster of sources in the magnitude--size plane, which the star selector identifies as stars. Unfortunately, the optics of the GOTO telescopes means that there is significant variation in the PSF across the focal plane, with point sources in the outskirts of the image being elongated and significantly larger than those in the central regions of the frame. We found that the star selector's default settings tended to exclude point sources in the outskirts of the image as it incorrectly classified them as being extended (and therefore excluded them for PSF determination; see top panel of fig. \ref{fig:psfSelection}). To overcome this restricting, we increased (i.e., relaxed) the range of source sizes that could be included as point sources by increasing the number of standard deviations before they are excluded from consideration (from 0.15 to 10) and by increasing the K-means sigma clipping threshold (from 2.0 to 10.0). These changes resulted in the inclusion of a more diverse range of object sizes which, as fig. \ref{fig:psfSelection} shows, improved the coverage of selected sources across the image.

With point sources selected, we used the LSST stack's own Principle Component Analysis (PCA; see, e.g., \citealt{Jee07}) module to model their PSF across the image. The fit is iterated, with each iteration rejecting outlying sources from the fit. What is considered an outlier is controlled via various adjustable parameters. Again, due to the complex and strongly spatially-varying nature of the GOTO PSF, we had to change a number of parameters that control the PSF modelling from their default values. Firstly, we increased the number of Eigen components from four to six, which helps to model the somewhat complex PSF in the outskirts of the GOTO frames. We found we also needed to increase (i.e., relax) the thresholds that the LSST stack uses to identify outliers. We increased the reduced-$\chi^2$ threshold above which sources are considered outliers and subsequently rejected from the next iteration, and the standard deviation threshold for rejecting sources from the the spatial fit (in both cases we set these parameters to 50). These adjustments resulted in significant improvements in the PSF model across the GOTO images, examples of which are shown in fig. \ref{fig:psfs}.

After modelling the PSF, we perform cosmic ray detection and repair using the LSST stack's built-in module based on the approach outlined in \cite{Bosch18}. When using the default parameters, we found that a number of stars in our image were being flagged as cosmic rays, resulting in their cores being inadvertently ``repaired''. As suggested in \cite{Bosch18}, we rectified this by reducing the \texttt{charImage.repair.cosmicray.cond3\_fac2} parameter within the \texttt{processCcd.py} config file to 0.1.

\subsubsection{Image calibration}
With the background subtracted and PSF characterised, we next used the LSST stack's own calibration modules to first find an astrometric solution for our science exposures. It then uses this solution to match detected sources to a catalogue of photometric standard stars to perform photometric calibration.

In terms of astrometry, each GOTO exposure includes in its header the requested RA and Dec of the mount pointing. The pointing of each individual UT, however, is typically offset by a number of degrees from this position. As such, while the RA and Dec included in the header can be used to provide a rough guide to the general region of sky covered by the image, it cannot be used to provide an accurate WCS solution. Instead, we used the \texttt{astrometry\_dot\_net} modules included in the LSST stack to obtain an accurate WCS for each image. Since the header astrometry of the incoming exposures is only accurate to within a number of degrees, we found we had to alter the LSST stack's default parameters considerably to reliably obtain an accurate WCS solution. While we used the requested mount pointing as a guide, we specified that the solver should search for a solution within 5 degrees of this position to account for the large potential offset between the mount pointing and the true pointing of the UT.

To find a WCS solution, the LSST stack runs a source detection algorithm (we used a high detection threshold of 30$\sigma$) and attempts to match detected sources to an astrometric reference catalogue. In our case, we used the UCAC4 catalogue (\citealt{Zacharias13}) as a reference as it is complete to $m_{\rm R}=16$, which is well-matched to the brightness of the high-significance sources in GOTO frames that we use for astrometric calibration. Again, since GOTO's raw image headers only provide a vague estimate of the true central RA and Dec of each image, we used a large matching radius of 120 ~arcsec to match between detections and the reference catalogue. As well as calculating the central RA and Dec, the LSST stack (in our case, via \texttt{astrometry\_dot\_net}) will also provide a full WCS solution, including Simple Image Polynomial (SIP) terms. Since we know the pixel scale of the UTs well, we considered only a 10\% uncertainty on this value to account for possible distortions, and used a third-order polynomial to fit any distortions. Using these parameters, we obtained a WCS solution with sub-arcsecond scatter for all our science frames, which is considerably smaller than the GOTO pixel scale of 1.24~arcsec per pixel (see section \ref{sec:res:ast}).

With a WCS solution found, the LSST stack is able to positionally-match detected sources to a photometric reference catalogue. We used the PanSTARRS Data Release 1 catalogue (\citealt{Chambers16}) as our photometric reference catalogue as it covers the same region of sky as covered by GOTO. However, since PanSTARRS is far deeper than GOTO's single-exposure detection limit, we filtered the catalogue to only include sources that are brighter than 19th magnitude, thereby reducing the size of the catalogue by roughly 90\%. We used a matching radius of 1 pixel (i.e., 1.24~arcsec) when matching to this photometric reference catalogue since our astrometry solution is typically accurate to sub-pixel scales.

Since GOTO's L-band does not match any of the PanSTARRS filters, we applied colour terms to the reference sources to convert the PanSTARRS magnitudes into (calibrated) synthetic GOTO magnitudes. To calculate the colour terms, we passed the \cite{Pickles98} catalogue of synthetic stellar spectral models through synthetic PanSTARRS and GOTO filter passbands. We chose the PanSTARRS g-band filter as a ``primary'' filter (i.e., the one that we felt most closely matches GOTO's L-band), and the PanSTARRS r-band filter as a ``secondary'' filter; together, these two filters encapsulate colour information on each source. Next, we generated a plot of $m^{\rm GOTO}_{L}-m^{\rm PS}_{\rm g}$ vs. $m^{\rm PS}_{\rm g}-m^{\rm PS}_{\rm r}$ containing points for all our spectral models. We then fit the resulting locus with a second-order polynomial. By feeding the LSST stack these coefficients via the \texttt{calibrate.photoCal.colorterms} config parameter contained in the \texttt{processCcd.py} config file, it is therefore able to calculate predicted L-band magnitudes from the g-band magnitudes and $g-r$ colours of sources in the PanSTARRS catalogue. Of course, these colour terms are only viable within the region of colour space covered by the spectra used to generate the synthetic PanSTARRS and GOTO magnitudes. We therefore limited the choice of reference stars by imposing appropriate colour limits during reference selection (i.e, $g-r>0$ and $r-i<0.5$). As we used PanSTARRS AB magnitudes, the GOTO resulting GOTO magnitudes are also in the AB magnitude system.
\\

\noindent
In total, the ISR, characterisation and calibration steps described in this section and performed by \texttt{singleFramedriver.py} took a total of 210~s per image, on average.

\subsection{Image coaddition}
\label{subsec:imageCoaddition}
After calibrating our individual science exposures, we next used the LSST stack to coadd these exposures to generate set of deep images from which we obtained a GOTO reference catalogue to be used as the basis of our forced photometry.
 
Prior to coaddition, the LSST stack reprojects each individual exposure onto a sky map which in our case is independent of GOTO's survey grid mentioned in section \ref{sec:goto}. This reprojection means that the coadds are unaffected by pointing errors (whereby repeat observations of the same tile aren't perfectly aligned) or reconfiguration of the full GOTO field-of-view due to the addition/removal of UTs. We chose to generate our sky map using HEALPix. Following the HEALPix model, the LSST stack divides each of the 12 base HEALPix pixels into $2^n$ ``tracts'' which are, in turn, split into ``patches''. In our case, we generated a whole sky map (using \texttt{makeSkyMap.py}), consisting of 192 (i.e., $n=4$) tracts containing patches of inner dimension of 4000$\times$4000 1.24~arcsecond pixels. We also specified a patch border of 100 pixels and a tract overlap of 0.1 degrees to ensure that there is overlap between both patches and tracts. We used a TAN projection and rotate the patches by 45 degrees to aid tessellation.

To perform the reprojection and coaddition of our individual exposures, we used the \texttt{coaddDriver.py} pipe driver, keeping most of its configuration parameters set to their default values. We do, however, turn off source detection, since that is performed in the next step. { In total, warping a single image, splitting it up into its individual patches, then coadding those patches took, on average, 440~s.}

\subsubsection{Source detection and deblending on coadded frames}
With coadded images generated, we next performed source detection and measurement on the coadds, with the resulting source catalogue used as a basis for forced photometry (described in Makrygianni et al. 2020). To perform source detection, we used the LSST stack's \texttt{multiBandDriver.py} task. As its name suggests, \texttt{multiBandDriver.py} is designed to both detect, measure, and match sources across multiple bands (see \citealt{Bosch18} for further details). At present, however, GOTO's main survey is only being performed in the L-band, so we used \texttt{multiBandDriver.py} to perform source detection and measurement in that single band and did not use its cross-band matching features. For detection on coadds, we largely used \texttt{multiBandDriver.py}'s default parameters, which correspond to a 5$\sigma$ detection threshold relative to the local noise. 

A major challenge facing all imaging surveys, especially those that cover crowded fields, is that of deblending detections into multiple sources. Problems with deblending culminate in either the failure to successfully separate multiple close objects (i.e, under-deblend), or deblend an individual object into multiple sources (i.e., over-deblend). Usually, the optimal outcome is a compromise between these two extremes.

\begin{figure}
    \includegraphics[width=\columnwidth]{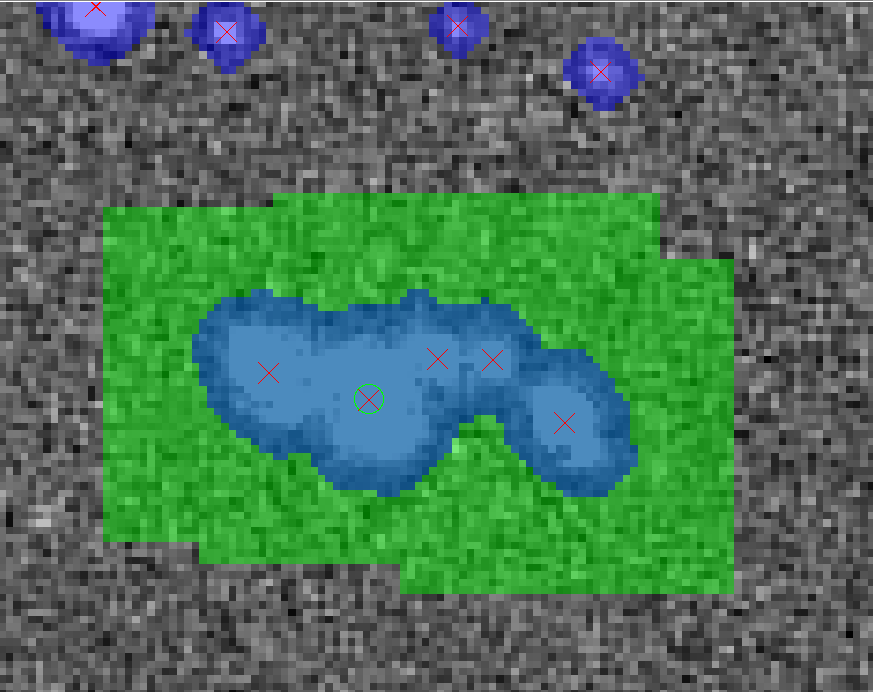}
    \caption{Deblending of a five-object detection. The detected pixels are masked with blue. It is clear that the detected {\fontfamily{qcr}\selectfont footprint} covers all the five sources and it is the deblender who will isolate the individual sources. The green circle shows the centroid of the parent source and the red crosses show the centroid of the {\fontfamily{qcr}\selectfont children} sources after deblending. The green shaded regions indicate the regions around where the fake sources have been injected.}
    \label{fig:blendedObject}
\end{figure}

We used the LSST stack's \texttt{meas\_deblender} package to deblend the detected sources, leaving most parameters at their default values aside from \texttt{maxFootprintArea}, which we reduced to 10,000 from $10^6$ as the latter caused the deblender to crash due to memory limits.\footnote{ At the time we started this project, \texttt{meas\_deblender} was the LSST stack's default deblender, but this has now been retired in preference of the \texttt{Scarlet} package (\citealt{Melchior18}).} Unfortunately, this change means that the deblender is unable to consider sources that cover very large numbers of pixels, such as very local galaxies (e.g., M31). Thankfully, such large sources are extremely rare and their study is not a high priority for GOTO science, so we felt that it is an acceptable loss to bear. We assessed the performance of the LSST deblender on GOTO images using artificially-inserted sources. Figure~\ref{fig:blendedObject} shows an example of a ``blended object'' consisting of five sources of different magnitudes artificially inserted into a coadd image. The five red crosses shows the positions of where the deblender has identified the centroids of the deblended sources. In such cases, the final catalogue of sources contains properties of both the parent (i.e., undeblended) source and the child (i.e., deblended) sources.

\begin{figure*}
    \includegraphics[width=\linewidth]{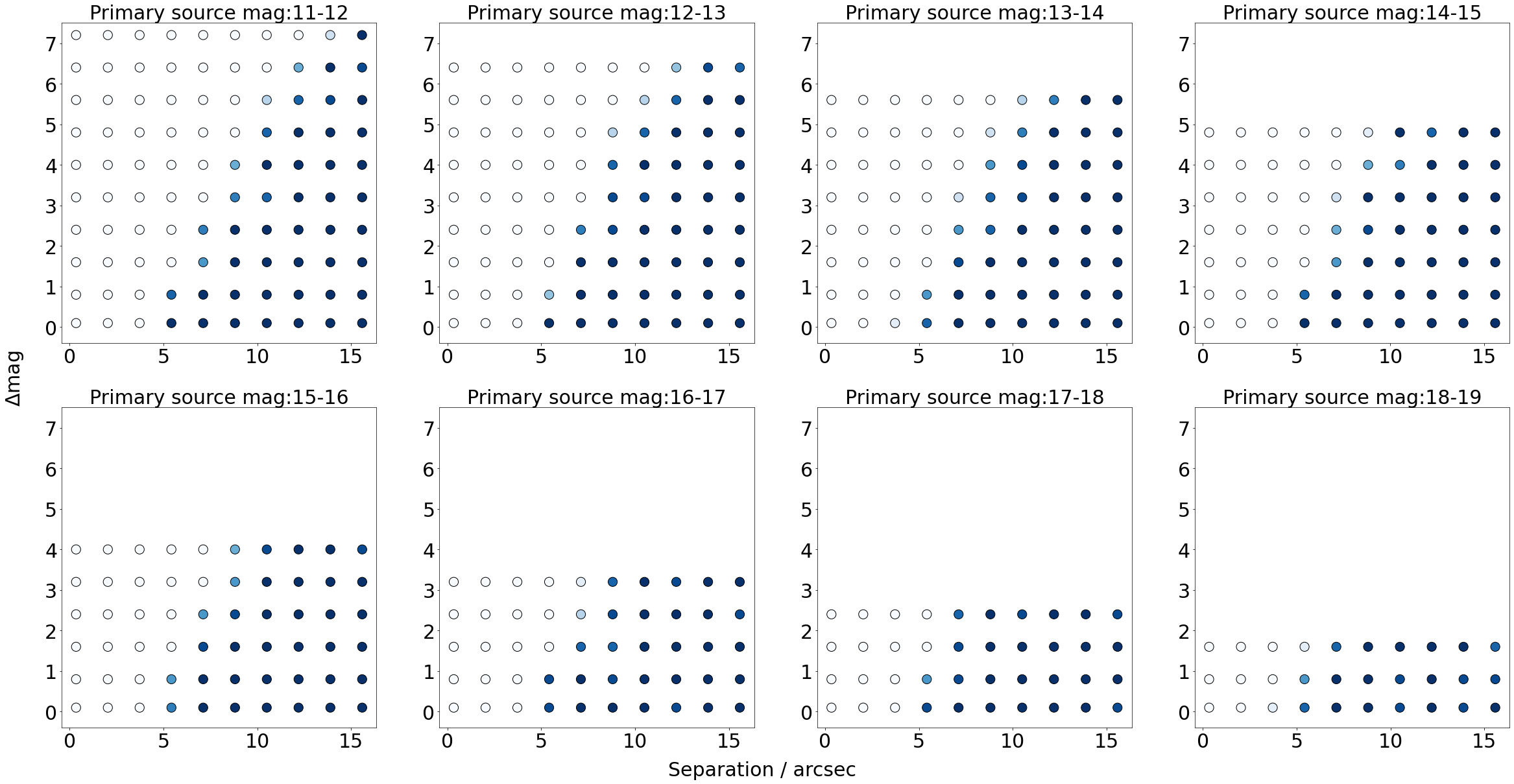}
    \includegraphics[width=\linewidth]{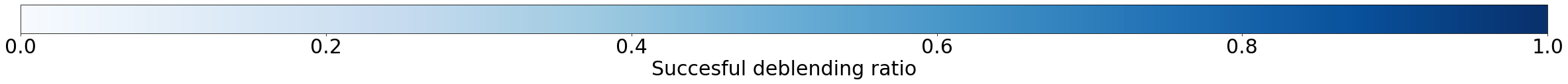}
	\caption{Plot showing the performance of the deblender to separate two injected (i.e., fake) sources across a range of separations (x-axis) and magnitude difference (y-axis). Each individual plot corresponds to a different ``primary'' source brightness, with the primary source always the brighter of the two. The colour of each point shows the fraction of ten pairs -- located at various points around the image -- that were successfully deblended (see colour bar). Overall, we find that the deblender is able to successfully separate sources that are 6~arcsec apart when those sources are of similar brightness, rising to 15~arcsec when they differ by $\sim7$~mag.}
    \label{fig:debtest2}
\end{figure*}

To obtain a quantitative assessment of the regions of the parameter space where the deblender breaks down on GOTO images, we injected 100 pairs of sources into one of our final, coadded images. We varied the separation of the pairs and their magnitude difference in order to assess how these parameters affect the success of the deblender. We performed this test ten times at ten locations across the coadded image. By using a range of combinations of separation and magnitude difference we obtained an estimate of the success rate of the deblender as a function of these properties. Fig. \ref{fig:debtest2} shows the results of this test. Each point corresponds to a different combination of separation/magnitude-difference, and is coloured according to the fraction of successful deblends. As one would expect, the deblender is more successful at deblending sources that are closer in brightness (i.e., small magnitude difference) and which are more widely separated. In the case of sources of similar magnitude, we find that 6~arcsec (i.e., $\approx$5~pixels) is the smallest separation we can deblend with close to approaching 100\% success rate, increasing to 15~arcsec when the sources differ in brightness by seven magnitudes.


One of the cases where the deblender performs poorly for GOTO images is in the case of very bright point sources, which are sometimes over-deblended (i.e., a single true source is split into multiple sources by the deblender). Further investigation of such over-deblended point sources showed that parent source is usually a saturated source. We choose to keep such cases of saturated sources in our final catalogues as they can easily be filtered-out using the \texttt{pixelFlag\_saturatedCenter} pixel flag. 

After deblending, \texttt{multiBandDriver.py} measures various types of source photometry. At this stage, our priority was to obtain a GOTO reference catalogue for forced photometry, which doesn't require a wide range of different types of photometry measurements. Therefore, to reduce processing time, we only measured circular aperture and Kron (\citealt{Kron80}) photometries on the coadded images. For the former, we used the following aperture radii: 3.72, 5.58, 7.44, 11.16, 14.88, 29.76, and 59.52~arcsec (corresponding to 3, 4.5, 6, 9, 12, 24, and 48 pixels, respectively). We also attempted to measure PSF photometry on the coadded images, but while this tends to work well on individual science frames (see Makrygianni et al. 2020), we found it delivered poor results on coadded frames. We suspect that the ability to obtain reliable PSF photometry on individual frames but not on coadds is related to the increased complexity of PSF models for the coadded frames, which are constructed from the weighted mean of the spatially-varying PSFs of each input image. If this is, indeed, the case then PSF photometry should become more reliable with improved PSF modelling and/or the exclusion of frames with poorer seeing when constructing the coadds.

In total, detection, deblending and measurement on coadded sources as described in this subsection took a total of 60~s per patch, on average.

\section{Results}

After running \texttt{multiBandDriver.py} on the coadded frames, the LSST stack detected and measured a total of 166 million deblended sources within the region of the sky covered by GOTO between the dates of 2019-02-24 and 2019-03-12 (see Fig. \ref{fig:coverage}). In this section, we assess the quality of these measurements. By comparing to external catalogues, we first assess the accuracy and precision of the astrometric measurements of the sources, which is particularly important for future forced photometry measurements. Next, we assess the quality of the colour-corrected photometry measurements, by again comparing to external reference catalogues. Finally, we calculate and report the depth of the catalogue obtained from the coadded images. Throughout this section, it is worth bearing in mind that both the LSST stack and {\sc gotophoto} are still under active development, and thus subsequent updates will likely lead to improvements in the astrometric and photometric measurements of both pipelines.

\subsection{Astrometry assessment}

To assess the quality of the astrometric measurements obtained by processing GOTO data with the LSST stack we performed a positional match to the PanSTARRS DR1 database (hereafter, PS1; \citealt{Flewelling16}), using a 1.24~arcsecond (i.e., 1 pixel) matching radius. PS1 has a standard deviation of the mean residuals of 2.3~mas in RA and 1.7~mas in Dec when compared to Gaia DR1 (\citealt{Magnier16}). As such, PS1 is sufficiently precise for our purposes. With our database of LSST stack-measured GOTO sources containing over 100 million sources, rather than positionally match our entire catalogue we instead randomly selected one out of every 50 sources to match to.

To avoid matching to artefacts or other defects we did not consider GOTO sources that are flagged as containing interpolated pixels. Similarly, sources that are flagged as containing saturated pixels were also excluded. Further, the final catalogue only consists of sources that are flagged as {\texttt detect\_isPrimary} which ensures that the catalogue doesn't include duplicates arising due to either deblending (since the original, undeblended source is also recorded in the output catalogue) or due to being detected in overlapping patches/tracts. To ensure that out results are not adversely affected by low number statistics we excluded the nine of our 113 tracts that have fewer than 2000 sources {\it after} the 1/50 selection (i.e., we only include tracts that contain more than 100,000 sources prior to our down-sampling). 

After performing the match to PS1 DR1 we found that the mean angular distance between the measured positions of GOTO detections and PS1 DR1 positions is 0.31" with a standard deviation is 0.23". Typically, the mean angular distance between the GOTO detections and the PS1 matches is less than half GOTO's pixel scale (i.e., 1.24~arcsec). Since we used the UCAC4 catalogue, rather than PS1, as an astrometric reference catalogue, it is possible that some of the aforementioned differences arise due to systematic differences between UCAC4 and PanSTARRS astrometry. A cross match between UCAC4 and PS1 for the same sources as used for the GOTO-PS1 comparison gives a mean positional difference of 0.27~arcsec, with a standard deviation of 0.20~arcsec. As such, differences between UCAC4 and PS1 likely account for a non-negligible portion of the discrepancy between the measured GOTO and PS1 positions. Astrometry could be improved, therefore, by adopting a more precise astrometric reference catalogue such as, for example, GAIA (\citealt{Gaia18b}). This would, however, require a major reprocessing of the data which we felt was not justified, considering that our astrometry solution is already good to sub-pixel levels.

\label{sec:res:ast}
\begin{figure}
    \includegraphics[width=\columnwidth]{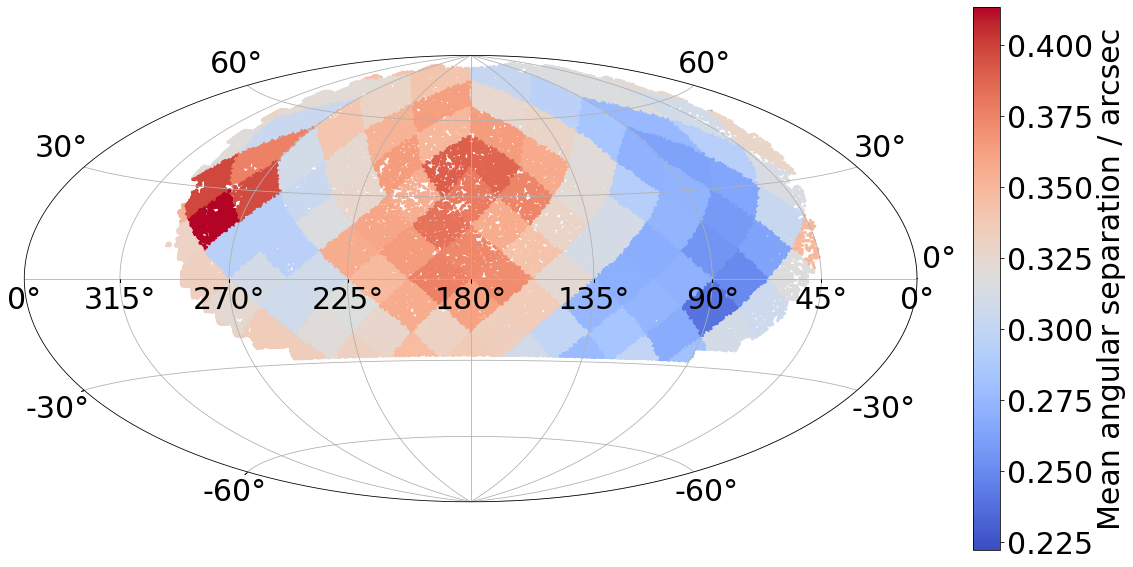}
    \centering
    \caption{Plot showing how the average astrometric offset between LSST stack-detected sources differs from matched PanSTARRS DR1 (PS1) sources across our full survey, on a per-tract basis. Each point represents a detected source (randomly down-sampled by a factor of 10), coloured according to the average astrometric offset of its tract.}
    \label{fig:cmapsAstrometry}
\end{figure}

Figure ~\ref{fig:cmapsAstrometry} shows the mean difference between GOTO and PS1 positions for each of our tracts. The three panels show the mean total positional difference, and the difference in right ascension and declination. It is clear from this figure that some tracts have worse astrometry than others. After some investigation, we determined that this is related to the quality, especially in terms of seeing, of the images that go into each coadd. As we mentioned in section 3, we produced the coadd images from all images obtained between the nights of the 2019-02-24 to the 2019-03-12 and did not filter on image quality (e.g. seeing). When producing future versions of the catalogue, we intend to select for seeing, which should reduce this inter-tract differences in astrometric precision.

\begin{figure}
    \includegraphics[width=\columnwidth]{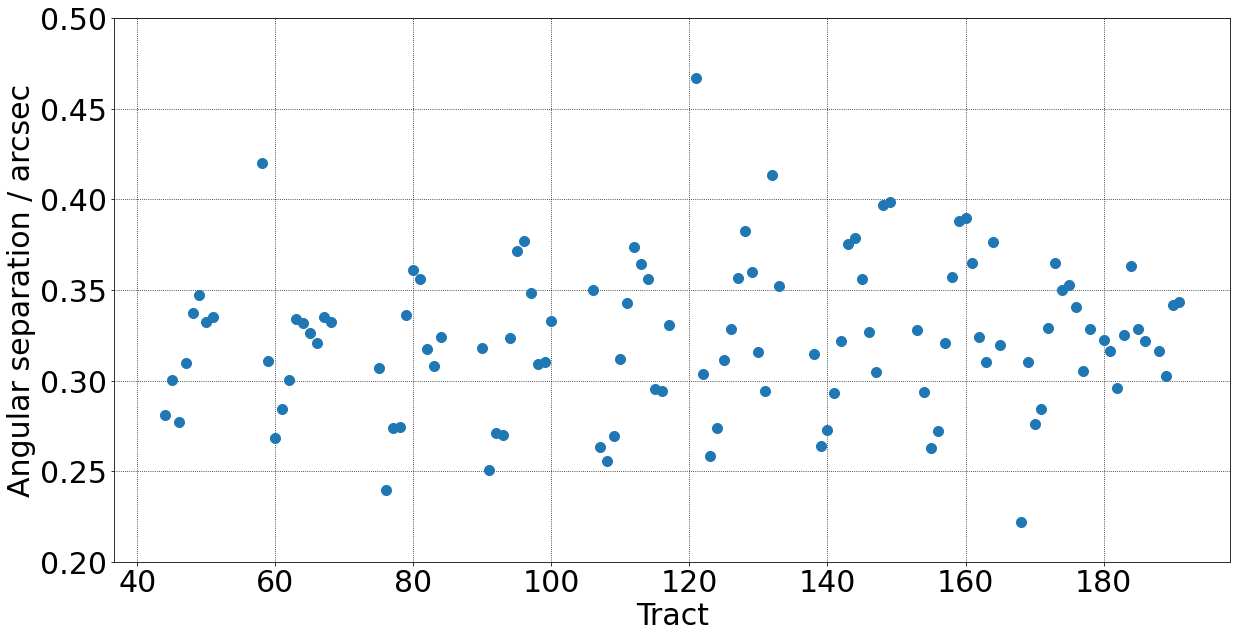}
    \centering
    \caption{The variations of the mean offsets between GOTO and PS1 positions as a function of the tract number. There is a repeating pattern roughly every 20 tracts which is due to the projection of the changes in average angular distance shown in Fig. \protect\ref{fig:cmapsAstrometry} onto the tract numbering system.}
    \label{fig:aDvsTract}
\end{figure}

As we described in section 3.1, the LSST stack software splits the sky into ``tracts'' and ``patches''. In Fig. ~\ref{fig:aDvsTract} we explore how the mean angular separation between matched sources changes as a function of tract number. In general, we find that the mean difference in astrometry is between 0.3~arcsec and 0.4~arcsec for each tract, although some have mean differences that are as high as 0.6~arcsec. The apparent sinusoidal pattern between mean astrometric difference and tract number is due to the projection of the positional dependence seen in Fig.~\ref{fig:cmapsAstrometry} onto the tract map.

To summarise, we emphasise that the typical astrometric offsets within GOTO data as measured by the LSST stack are small compared to GOTO's pixel scale, and the astrometry is sufficiently precise and accurate to match to other optical and/or near-infrared catalogues.

\subsection{Photometry assessment}

\begin{figure*}
\includegraphics[width=\columnwidth]{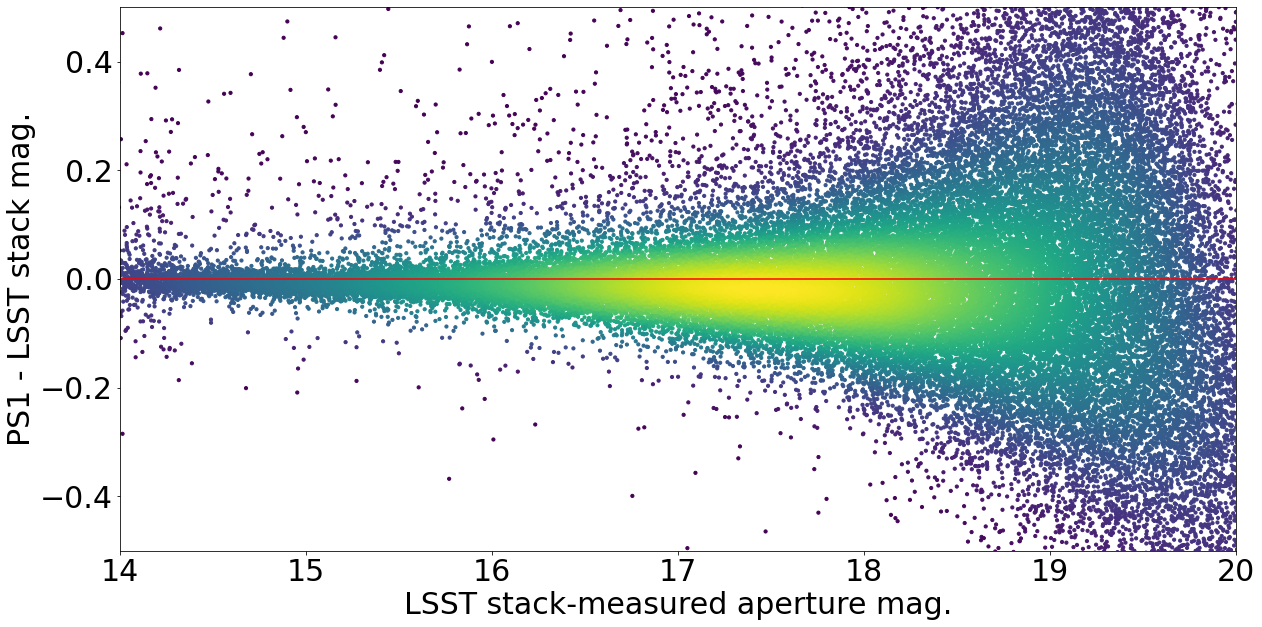}
    \includegraphics[width=\columnwidth]{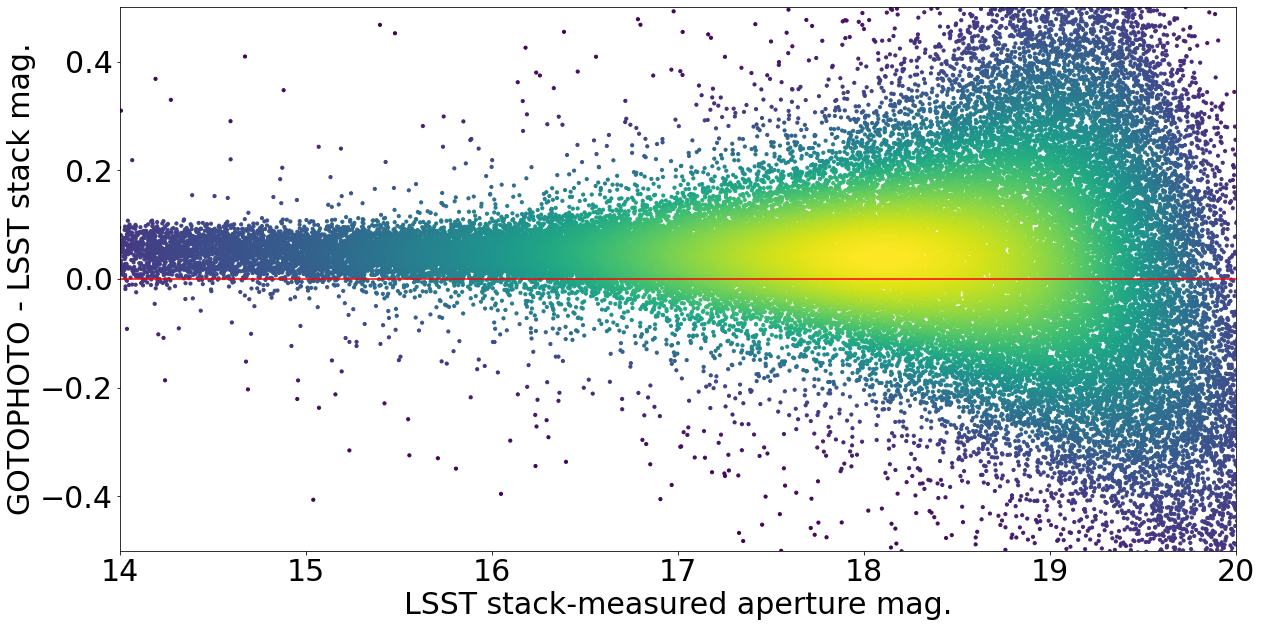}
    \centering
    \caption{Plots showing the difference in measured photometry between LSST stack-detected sources (measured in 9 pixel radius apertures) and positionally-matched PS1 (left) and {\sc gotophoto}-detected sources (right), as a function of LSST stack-measured magnitude. These plots show sources within a single tract (tract 122), but all other tracts show similar results. The PS1 photometry has been colour-corrected so that it corresponds to GOTO's L-band. {\sc gotophoto} does not yet colour-correct its reference magnitudes when calculating zero-points, hence the observed offset. In the case of sources brighter than $m_{\rm L} \sim16$, LSST stack-measured photometry is consistent with PS1 photometry to within 50~mmag. We find, however, that there are fewer catastrophic outliers when comparing to the {\sc gotophoto}-measured photometries. This suggests that a significant number of catastrophic outliers in the left-hand plot are intrinsic to the observations (e.g., variable sources), rather than due to problems with the pipeline, since {\sc gotophoto} processed the same data.}
    \label{fig:coaddPS}
\end{figure*}

In this subsection we assess the quality of the aperture photometry of GOTO sources detected and measured by the LSST stack. We do this by comparing to positionally-matched PS1 sources (see Section \ref{sec:res:ast}) and sources measured by GOTO's in-house pipeline, {\sc gotophoto}, which uses SExtractor (\citealt{Bertin96}) to detect and measure sources. Starting with the comparison against PS1, we show in the left-hand plot of Fig.~\ref{fig:coaddPS} the magnitude difference between (colour-corrected) PS1 PSF photometry and GOTO photometry (using 11.16~arcsec apertures) as a function of the LSST stack-measured source magnitude. In this case, we have only plotted the results for a single tract (i.e., tract 122); we explore how the photometric quality changes as a function of tract later in this section. From Fig.~\ref{fig:coaddPS}, we see that for sources brighter than $m_{\rm L}\sim18$ the LSST stack-measured GOTO photometry is typically within 0.2 magnitudes of the colour-corrected PS1 photometry, while sources brighter than $m_{\rm L}\sim16$ are typically well within 50 mmag of the colour-corrected PS1 magnitudes.

Next, we compare the GOTO photometry measured using the LSST stack against that measured using {\sc gotophoto}, which is also calibrated using PS1 photometry. This is a useful comparison as in doing so we are comparing data obtained using the same system but processed using different pipelines. There are, however, some details that we need to take into account while assessing the results of this comparison. As mentioned in section 3.1.4, when photometrically calibrating the GOTO exposures using the LSST stack, we used colour-corrected PS1 magnitudes as a reference. By contrast, {\sc gotophoto} did not apply a colour correction to the PS1 magnitudes. When performing the comparison, we chose measurements from the {\sc gotophoto} database that had been made on the same exposures as those used to create our (LSST stack-produced) coadded images. We present the results of our comparison between LSST-stack and {\sc gotophoto}-measured photometry in the right-hand panel of Fig.~\ref{fig:coaddPS}, again plotting magnitude difference against LSST-stack measured magnitude. We see a systematic offset between the two photometry measurements of around 0.1 magnitudes, plus a broader scatter at brighter magnitudes than that seen when compared to the PS1 comparison. We attribute both to the lack of colour-correction in the {\sc gotophoto} data. However, we see far fewer catastrophic outliers in the {\sc gotophoto} comparison, which suggests the outliers in the PS1 comparison are intrinsic to the exposures (e.g., variable sources, defects, etc.).

\begin{figure}
    \includegraphics[width=\columnwidth]{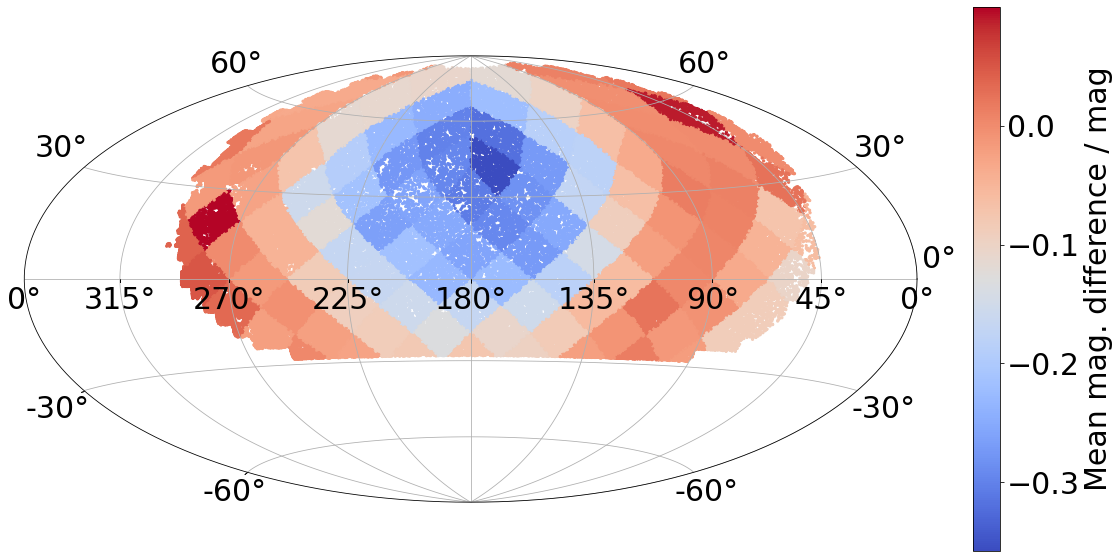}
    \includegraphics[width=\columnwidth]{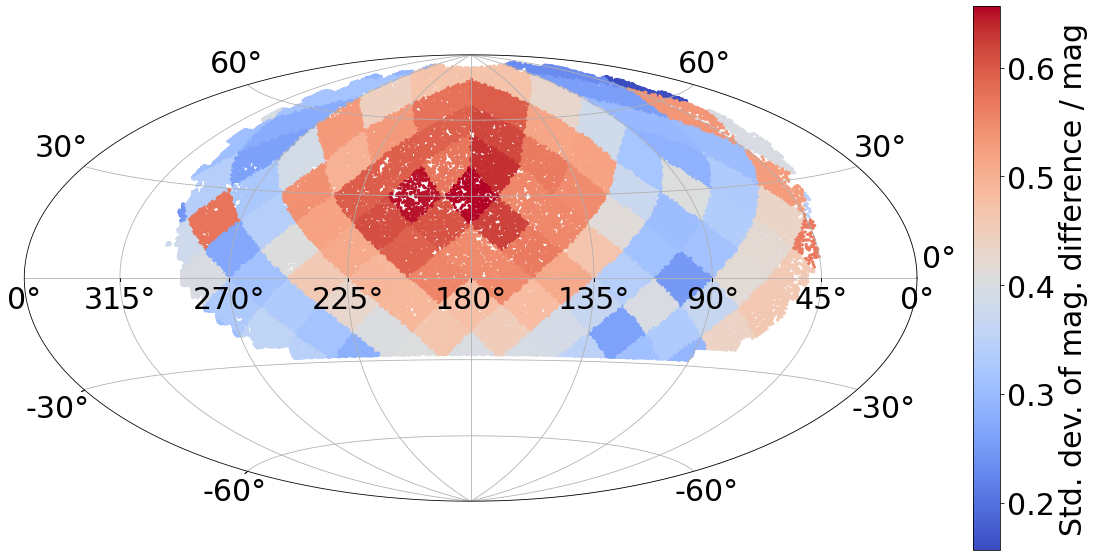}
    \centering
    \caption{Mean (top) and standard deviation (bottom) of the difference between the 9~arcsec aperture magnitudes as measured by the LSST stack on the coadded frames and the colour-corrected PS1 magnitudes, on a per-tract basis. Each point represents a detected source (randomly down-sampled by a factor of 10), coloured according to the mean difference or standard deviation of its tract.}
    \label{fig:coaddPSmap}
\end{figure}

After assessing the photometric quality of a single tract, we now explore how the photometric quality changes as a function of sky position. As we are now considering all tracts, we compare against the same PS1-matched catalog as used for the astrometric assessment described in the previous section. To do this, we plot summary statistics of photometric quality for sources across footprint covered by our coadded images. In the left-hand panel of Fig. \ref{fig:coaddPSmap} we plot the mean difference between LSST-measured GOTO photometry and PS1 photometry for each of our tracts, whereas in the right-hand panel we show the standard deviation of this difference. From the left-hand panel of fig.~(\ref{fig:coaddPSmap}) we see that some of the tracts centred around (RA,\ Dec)=(175, 30) show significantly larger mean differences and also larger standard deviations than other tracts. These coincide with the regions of the sky which show evidence of poorer astrometry, leading us to believe that the two are due to the same underlying cause: the poorer quality of the images that went into producing these coadds.

\subsection{Survey depth and detection completeness}

Having assessed the quality of the photometric measurements conducted by the LSST stack on GOTO coadd images, we now use these measurements to estimate the limiting sensitivity (i.e., depth) of the coadd survey. We use two methods to assess the depth of the coadd survey. Our first approach is to calculate the average magnitude of a source detected at a significance of 5$\sigma$. Our second approach is via injecting fake sources of varying brightness into the coadded images and computing the fraction of recovered sources as a function of brightness. Both approaches, together with their results, are described in more detail below.

\begin{figure}
    \includegraphics[width=\columnwidth]{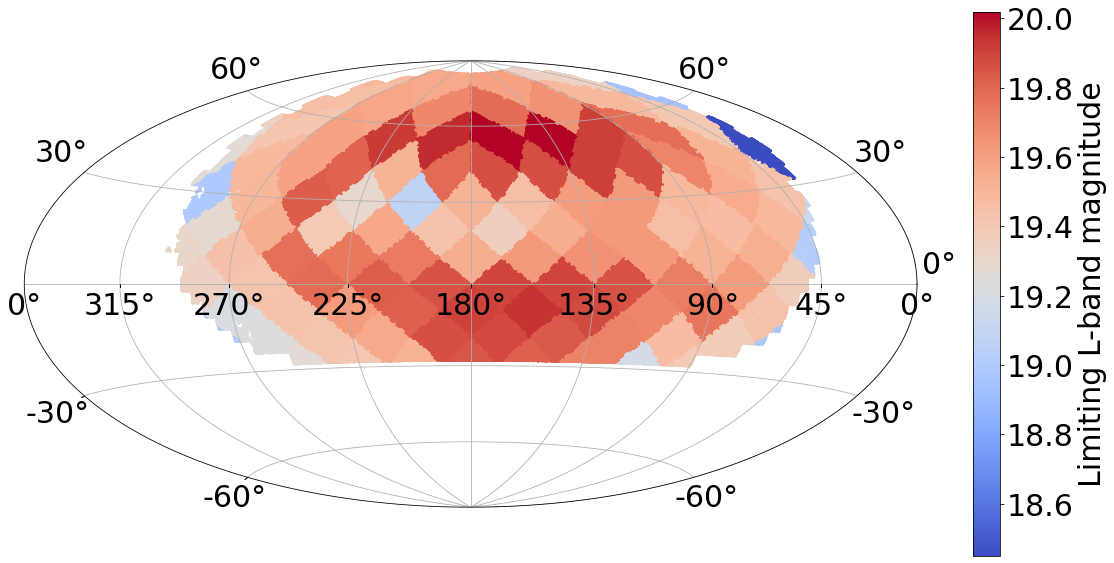}
    \centering
   \caption{The depth for each tract in the coadded survey. The colour bar on the right corresponds to a $5\sigma$ detection. Overall, we report a current L-band survey depth of 19.6~mag, which is set to increase with GOTO's repeated observations of the sky.}
    \label{fig:surveydepth}
\end{figure}

To estimate the 5$\sigma$ limiting magnitudes of our coadded images we computed the mean aperture magnitude of sources with S/N values between 4.5 and 5.5. Following \cite{Aihara18}, we only use sources flagged as {\texttt detect\_isPrimary~=~True} which filters duplicate sources from the catalogue (see Section \ref{sec:res:ast}). We also exclude sources that are flagged as containing either interpolated pixels (largely due to defects) or saturated pixels. Figure~\ref{fig:surveydepth} shows how the depth measured in this way varies across a single tract. Since the coadd survey is made up of many different exposures taken on different nights in different conditions, the survey depth varies between coadded images (and hence patches and tracts) depending on the quality of the individual images that went into the building of a specific coadd. Again, this variation could be reduced by selecting according to conditions (i.e., photometric quality and seeing). In general, however, the survey reaches an L-band depth of $\sim$19.6, as assessed by this method. Recall, however, that this is based on coadds made up of a ``single pass'' of GOTO observations consisting of three 1-minute back-to-back observations. This depth will, of course, increase with as more exposures are added to the coadds.    

\begin{figure}
    \includegraphics[width=\columnwidth]{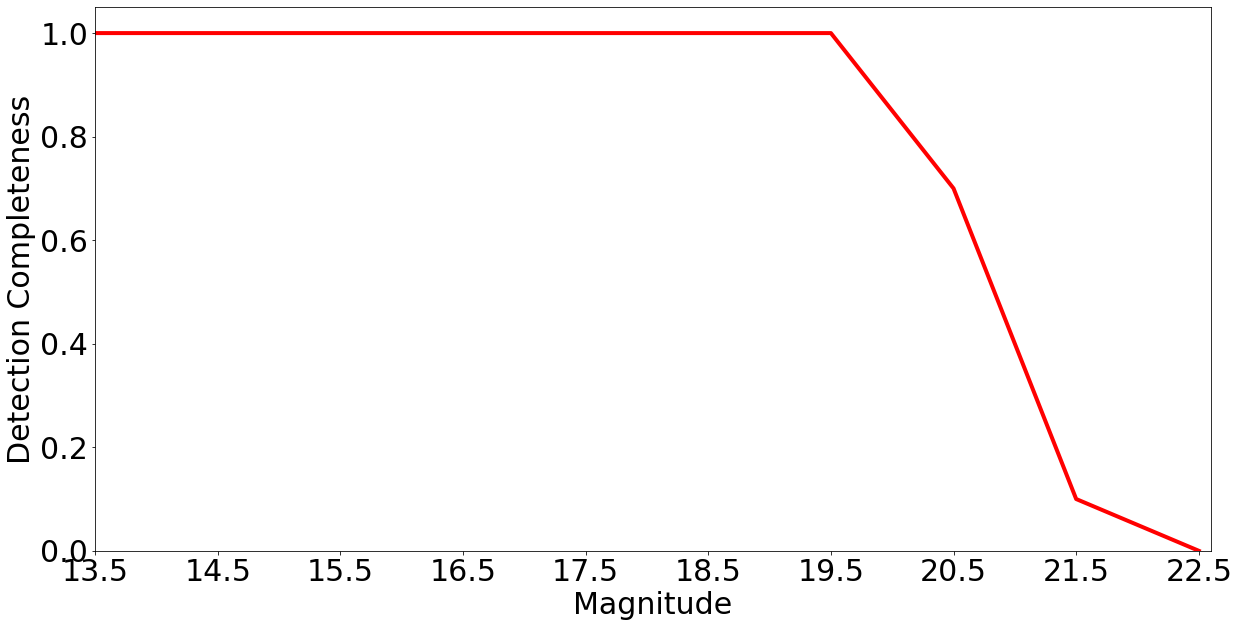}
    \caption{L-band detection completeness as a function of L-band magnitude in tract 94, patch 5,8 using artificial sources. This plot shows the fraction of injected sources that are detected at a significance of $>5\sigma$ in bins of 0.5 magnitudes. }
    \label{fig:detcompl}
\end{figure}

To characterise the survey depth according to detection completeness, we inserted artificial point sources into the the coadd images prior to (re-)running the LSST stack's detection algorithm. We used the \path{lsst.synpipe.positionStarFakes.PositionStarFakesTask} module within in the LSST stack to insert the fake sources into the coadd at random positions across a patch. The inserted sources spanned a magnitude range of $13 \leq m_{\rm L} \leq 23$, split into ten intervals. The catalogue of injected sources was constructed to contain ten sources per magnitude interval. After running the detection algorithm we calculated the fraction of inserted sources detected by the LSST stack at $>5\sigma$ as a function of magnitude. We performed this analysis for ten different patches, all in different tracts. The results of this analysis for one patch (patch $[5,8]$ of tract 94) is shown in Fig.~\ref{fig:detcompl}, in which we plot the fraction of recovered sources (i.e., completeness) as a function of magnitude. In this patch, our injected-source analysis shows that survey is 100\%\ complete down to $m_{\rm L}\sim19.5$ and 50\% complete at $m_{\rm L}\sim21$ mag. We do not detect any sources fainter than $m_{\rm L}\sim22.5$ at $5\sigma$. In each of the ten patches, we find that the 100\%\ completeness level is broadly consistent with the $5\sigma$ depth of that patch as measured using our other approach based on real (i.e., not injected) sources.   

\section{Conclusions}
\label{sec:conc}
We have adapted the LSST stack to process data collected by the Gravitational-wave Optical Transient Observer (GOTO). The wide field of view and high cadence -- if not the depth -- of GOTO data bears many similarities to that expected to be delivered by the LSST. As such, we were motivated to explore whether the LSST stack could be used as a secondary pipeline to GOTO's main in-house pipeline, {\sc gotophoto}. While a non-negligible amount of adaptation is required, we have found that, overall, the LSST stack can be used to successfully process data from facilities such as GOTO.

The main focus of this paper has been to describe how we adapted the LSST stack -- via our {\texttt obs\_goto} package -- to process data obtained by GOTO up to the point of frame coaddition and source detection on the resulting coadds (see section \ref{sec:obs}). In a second, accompanying paper (Makrygianni et al. 2020), we describe how we then use the LSST stack to perform forced photometry measurements at the positions of sources within the coadds.

In addition to describing our adaptations to the LSST stack, we also performed a number of data quality assessments. These focussed on the quality of the astrometry and photometry as measured by the LSST stack. After comparing against the PS1 catalogue, we report a mean astrometric offset of roughly 0.3-0.4~arcsec across almost the whole sky area considered (a small region had mean offsets of around 0.6~arcsec). For comparison, GOTO's pixel scale is 1.24~arcsec per pixel, meaning the average astrometric offsets are consistently below one pixel-width. In terms of photometry, once colour-terms are included to account for the differences between the GOTO and PanSTARRS filters, we find that the LSST stack-reported aperture photometry measurements are typically within 50 mmag of the PS1-reported aperture photometry for sources brighter than $m_{\rm L}\sim16~{\rm mag}$,increasing to 0.2~mag for sources with $m_{\rm L}~18~{\rm mag}$. Finally, by using these calibrated fluxes, we estimate that the GOTO coadds we generate reach a $5\sigma$ depth of $m_{\rm L}\sim19.6$.

\begin{acknowledgements}
We thank the anonymous referee for their comments, which significantly improved and clarified key aspects regarding the LSST Science Pipelines. We also thank those members of the LSST Community Forum whose advice enabled us to develop our \texttt{obs\_goto} and \texttt{obs\_necam} packages. The Gravitational-wave Optical Transient Observer (GOTO) project acknowledges the support of the Monash-Warwick Alliance; Warwick University; Monash University; Sheffield University; the University of Leicester; Armagh Observatory \& Planetarium; the National Astronomical Research Institute of Thailand (NARIT); the University of Turku; Portsmouth University; and the Instituto de Astrof\'{i}sica de Canarias (IAC). This study also makes use of PanSTARRS data. The Pan-STARRS1 Surveys (PS1) and the PS1 public science archive have been made possible through contributions by the Institute for Astronomy, the University of Hawaii, the Pan-STARRS Project Office, the Max-Planck Society and its participating institutes, the Max Planck Institute for Astronomy, Heidelberg and the Max Planck Institute for Extraterrestrial Physics, Garching, The Johns Hopkins University, Durham University, the University of Edinburgh, the Queen's University Belfast, the Harvard-Smithsonian Center for Astrophysics, the Las Cumbres Observatory Global Telescope Network Incorporated, the National Central University of Taiwan, the Space Telescope Science Institute, the National Aeronautics and Space Administration under Grant No. NNX08AR22G issued through the Planetary Science Division of the NASA Science Mission Directorate, the National Science Foundation Grant No. AST-1238877, the University of Maryland, Eotvos Lorand University (ELTE), the Los Alamos National Laboratory, and the Gordon and Betty Moore Foundation. This study also makes use of software developed for the Vera C. Rubin Observatory project, which is supported in part by the National Science Foundation through Cooperative Agreement AST-1258333 and Cooperative Support Agreement AST1836783 managed by the Association of Universities for Research in Astronomy (AURA), and the Department of Energy under Contract No. DE-AC02-76SF00515 with the SLAC National Accelerator Laboratory.
\end{acknowledgements}

\begin{appendix}
\label{appendix}
\section{\texttt{obs\_necam}: A bare-bones LSST obs package}
To conduct this study, we had to develop our own \texttt{obs\_goto} package in order for the LSST stack to be able to process GOTO data. As described in the main text, an obs package provides the LSST stack with information about the detector, the location and format of input and output data, and controls the operation of the stack via a number of configuration files. All this information is specific to a given telescope/detector system, so any project wishing to use the LSST stack to process their data will need to develop their own obs package.

At the time of writing, there are a number of obs packages available via the Rubin Observatory's github pages (\url{https://github.com/lsst}). All of these, however, have been written for a particular telescope system, and so include large amounts of information that is highly specific to those systems. We felt, therefore, that providing a generic obs package that others could adapt for their own specific purposes may provide an easier route to developing new obs packages for other systems. { This generic obs package is called \texttt{obs\_necam} and supports both the Generation 2 and Generation 3 Butlers}. It is available, together with instructions for its use, on an open licence at \url{https://github.com/jrmullaney/obs_necam}.

\end{appendix}

%\bibliographystyle{pasa-mnras}
%\bibliography{gotowlsst}
\input{gotowlsst.bbl}

\end{document}